\documentclass[aps,showpacs,superscriptaddress,preprint]{revtex4}%
\usepackage{amsfonts}
\usepackage{amsmath}
\usepackage{amssymb}
\usepackage{graphicx}%
\begin{document}
\title{Excitonic condensation for the surface states of topological insulator bilayers}
\author{Zhigang Wang}
\affiliation{LCP, Institute of Applied Physics and Computational Mathematics, P.O. Box
8009, Beijing 100088, People's Republic of China}
\author{Ningning Hao}
\affiliation{Institute of Physics, The Chinese Academy of Sciences, Beijing 100080,
People's Republic of China}
\author{Zhen-Guo Fu}
\affiliation{State Key Laboratory for Superlattices and Microstructures, Institute of
Semiconductors, Chinese Academy of Sciences, P. O. Box 912, Beijing 100083,
People's Republic of China}
\affiliation{LCP, Institute of Applied Physics and Computational Mathematics, P.O. Box
8009, Beijing 100088, People's Republic of China}
\author{Ping Zhang}
\thanks{Corresponding author. Email address: zhang\_ping@iapcm.ac.cn}
\affiliation{LCP, Institute of Applied Physics and Computational Mathematics, P.O. Box
8009, Beijing 100088, People's Republic of China}


\begin{abstract}
We propose a generic topological insulator bilayer (TIB) system to study the
excitonic condensation with self-consistent mean-field (SCMF) theory. We show
that the TIB system presents the crossover behavior from the
Bardeen-Cooper-Schrieffer (BCS) limit to Bose-Einstein condensation (BEC)
limit. Moreover, by comparison with traditional semiconductor systems, we find
that for the present system the superfluid property in the BEC phase is more
sensitive to electron-hole density imbalance and the BCS phase is more robust.
Applying this TIB model into Bi$_{2}$Se$_{3}$-family material, we find that
the BEC phase is most probable to be observed in experiment. We also calculate
the critical temperature for Bi$_{2}$Se$_{3}$-family TIB system, which is
$\mathtt{\sim}100$ K. More interestingly, we can expect this relative
high-temperature excitonic condensation since our calculated SCMF critical
temperature is approximately equal to the Kosterlitz-Thouless transition temperature.

\end{abstract}
\maketitle

\section{Introduction}

Recent technological advances in microfabrication bring growing interests in
studying exciton condensation in different bilayer physical systems such as
the semiconductor electron-hole bilayers \cite{Snoke,Sahin,Zhu} and graphene
bilayers \cite{CHZhang,MacDonald}. A number of novel physical phenomena are
obtained in these systems, such as the BCS-BEC crossover \cite{Comte} as well
as the subtle phase transition in the crossover region induced by the density
imbalance \cite{Strinati}, the dark and bright excitonic condensation under
spin-orbit coupling \cite{Can}, anomalous exciton condensation in high Landau
levels in magnetic field \cite{MacDonald}, room-temperature superfluidity in
graphene bilayers \cite{Mac}, etc. The conventional electron-hole bilayers are
fabricated with semiconductor heterostructures such as GaAs/AlGaAs/GaAs. The
character of the semiconductor electron-hole bilayers is that the electron and
hole bands are quadratic ones with different effective masses, which means
missing particle-hole symmetry in these kinds of systems and small superfluid
density. Hence, in semiconductor electron-hole bilayers, the excitonic
condensation needs very low temperature. Another better candidate for
electron-hole bilayers is graphene, which has a two-dimensional (2D) massless
linear Dirac-band structure in low energy limit. However, the coupling between
different Dirac-cone structures in the same Brilliouin zone brings flaw to
graphene to fabricate electron-hole bilayers \cite{Franz}.

On the other hand, another growing interest in condensed matter physics is the
very recent theoretical prediction \cite{Bernevig} and experimental
verification \cite{Konig} of the topological insulators \cite{Kane} (TIs) with
strong spin-orbit interaction. Several three-dimensional (3D) solids, such as
Bi$_{1-x}$Sb$_{x}$ alloys, Bi$_{2}$Se$_{3}$-family crystals, have been
identified \cite{Fu,Hsieh,HJZhang,Xia,Chen} to be strong TIs possessing
anomalous band structures. The energy scale for the surface states of these 3D
TIs is dominated by the $k$-linear spin-orbit interaction. Especially, the
strong TIs surface has single Dirac-cone band structure which is also
different from graphene. As a result, it is expected that the excitonic
condensate of these topological surface states probably have new characters.
\begin{figure}[ptb]
\begin{center}
\includegraphics[width=0.6\linewidth]{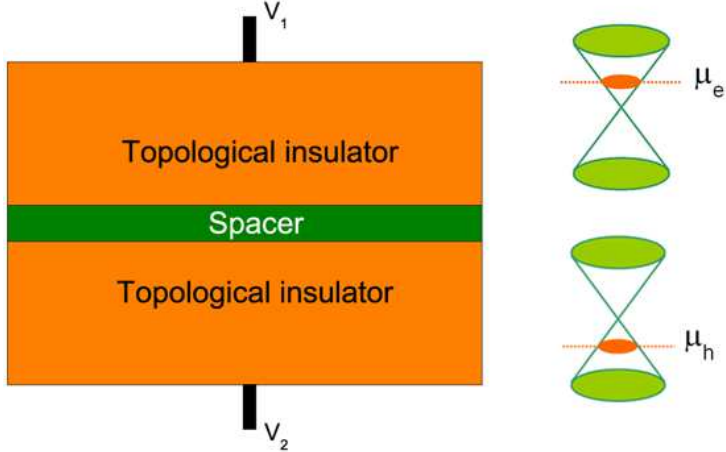}
\end{center}
\caption{(Color online) Left panel: Schematic structure of double-well
topological insulators in $x$-$y$ plane. The external gates can independently
tune the electron and hole densities. Right panel: The linear energy
dispersion around the Dirac point of the electrons and holes. }%
\label{f1}%
\end{figure}

Inspired by this expectation, in this paper we propose a topological insulator
bilayer (TIB) model analogous to Ref. \cite{Franz1}, a gated double TI layers
separated by an insulating spacer. Using this TIB model, we numerically study
the excitonic condensation of TI surface states. We find that the system also
presents BCS-BEC crossover along with the change in carrier densities in zero
temperature limit. However, there are two characters different from those of
conventional excitonic condensation in semiconductor bilayer systems. The
first is that the BCS phase of TIB is more robust than that of the
semiconductor bilayer systems; the second is that the superfluidity of the TIB
is more sensitive to the electron-hole density imbalance than that of the
semiconductor bilayer systems. These two characters physically root in the
$k$-linear band dispersion of the TIB. Moreover, by putting this TIB model in
Bi$_{2}$Se$_{3}$-family material, we investigate the excitonic condensation
and only find the BEC phase occurring due to the values of the parameters of
the material. The critical temperature of excitonic condensation in Bi$_{2}%
$Se$_{3}$-family TIB is also calculated in the self-consistent mean-field
(SCMF) approximation ($\sim$ $100$ K), which is found to be higher than that
in the traditional semiconductor electron-hole bilayers. More interestingly,
we can expect this relative high-temperature excitonic condensation since our
calculated SCMF critical temperature is approximately equal to the
Kosterlitz-Thouless (KT) transition temperature.

\section{The TIB Model}

The TIB system is schematically illustrated in the left panel in Fig.
\ref{f1}. Two TI films are separated by an insulating spacer of thickness $d$,
and the electron (hole) density can be independently tuned by the external
gate voltage $V_{1}$ ($V_{2}$). The linear dispersions of the TIs around Dirac
point are cartoonishly depicted in the right panel in Fig. \ref{f1}. The
grand-canonical Hamiltonian describing this TIB system can be written as
\begin{align}
H  &  =-\sum_{p,\mathbf{k,}\sigma}\mu_{p}\hat{p}_{\mathbf{k}\sigma}^{\dag}%
\hat{p}_{\mathbf{k}\sigma}+\sum_{p,\mathbf{k}}\hslash v_{F}^{p}\left(
k_{x}-ik_{y}\right)  \hat{p}_{\mathbf{k}\uparrow}^{\dag}\hat{p}_{\mathbf{k}%
\downarrow}+h.c.\nonumber\\
&  +\frac{1}{2\Omega}\sum_{p,p^{\prime}}\sum_{\mathbf{k},\mathbf{k}^{\prime
},\mathbf{q,}\sigma,\sigma^{\prime}}V_{\mathbf{q}}^{pp^{\prime}}\hat
{p}_{\mathbf{k}+\mathbf{q}\sigma}^{\dag}\hat{p}_{\mathbf{k}^{\prime
}-\mathbf{q}\sigma^{\prime}}^{\prime\dag}\hat{p}_{\mathbf{k}^{\prime}%
\sigma^{\prime}}^{\prime}\hat{p}_{\mathbf{k}\sigma}. \label{formula1}%
\end{align}
Here, $\mathbf{k}$, $\mathbf{k}^{\prime}$, and $\mathbf{q}$ are 2D wave
vectors in the layers, $\Omega$ is the quantization volume. $\mu_{p}$ is the
chemical potential for electron layer ($p$=$e$) or hole layer ($p$=$h$).
$\hat{p}_{\mathbf{k}\sigma}$ indicates the annihilation operator of electron
at the wave vector $\mathbf{k}$ and spin $\sigma$ (=$\uparrow,\downarrow$) for
electron layer ($p$=$e$), and hole layer ($p$=$h$). Note that $v_{F}^{e}%
$=$v_{F}$ and $v_{F}^{h}$=$-v_{F}.$The surface states of the strong TI film
have the linear dispersion: $\epsilon_{\mathbf{k}e,h}$=$\pm\hslash
v_{F}|\mathbf{k}|$. $V_{\mathbf{q}}^{pp^{\prime}}$ is the Fourier transform of
the Coulomb interaction: the intralayer Coulomb repulsive interaction
$V_{\mathbf{q}}^{ee}$($V_{\mathbf{q}}^{hh}$)=$2\pi e^{2}/\left(
q\varepsilon\right)  $, and the interlayer Coulomb attractive interaction
\cite{Balatsky,Shim} $V_{\mathbf{q}}^{eh}$=$-2\pi e^{2}\exp\left(  -qd\right)
/\left(  q\varepsilon\right)  $, which indicates that on the one hand, in the
limit of $d\rightarrow0$, the interaction between electron and hole becomes
that in monolayer; on the other hand, in the large thickness limit
$d\rightarrow\infty$, the interactions between the electrons in upper layer
and holes in lower layer should vanish. Here, $\varepsilon$ is the background
dielectric constant. Furthermore, for the present TIB system, the two TI films
are separated by an inulating spacer such as SiO$_{2}$, and the spin-orbit
interaction in the spacer is obviously negligible. Thus that it can be
expected that our model is appropriate in neglectering the interlayer hopping coupling.

In the basis $(\hat{e}_{\uparrow},\hat{e}_{\downarrow},\hat{h}_{\uparrow}%
,\hat{h}_{\downarrow})^{T}$, the Hamiltonian (\ref{formula1}) can be decoupled
to $H_{MF}$ under the mean-field approximation: $\Delta_{\sigma\sigma^{\prime
}}(\mathbf{k})$=$\sum_{\mathbf{q}}V^{eh}(\mathbf{q})\langle\hat{e}%
_{\mathbf{k}+\mathbf{q},\sigma}^{\dag}\hat{h}_{\mathbf{k}+\mathbf{q}%
,\sigma^{\prime}}\rangle$, $\Sigma_{\sigma\sigma^{\prime}}^{(p)}(\mathbf{k}%
)$=$-\sum_{\mathbf{q}}V^{pp}(\mathbf{q})\langle\hat{p}_{\mathbf{k}%
+\mathbf{q},\sigma}^{\dag}\hat{p}_{\mathbf{k}+\mathbf{q},\sigma^{\prime}%
}\rangle$. Then, $H_{MF}$ can be diagonalized with a 4$\times$4 unitary matirx
$U(\mathbf{k})$, $U^{\dag}(\mathbf{k})H_{MF}(\mathbf{k})U(\mathbf{k}%
)$=$\operatorname{diag}(E_{1}(\mathbf{k}),E_{2}(\mathbf{k}),E_{3}%
(\mathbf{k}),E_{4}(\mathbf{k}))$. The unitary matrix $U(\mathbf{k})$ is
construsted by the normalized eigenfunctions of the Hamiltonian
(\ref{formula1}), which can be numerically calculated by diagonalizing the
Hamiltonian matrix (\ref{formula1}) in the basis $(\hat{e},\hat{e},\hat
{h},\hat{h})^{T}$. Explicitly, the elements $U_{ij}(\mathbf{k})$ denotes the
$i$-th component of the eigenfunction corresponding to the eigenvalue $E_{j}$.
The relevant mean-field equations to be solved for the variables $\mu_{e}$,
$\mu_{h}$, and the gap functions $\Delta_{\sigma\sigma^{\prime}}(\mathbf{k})$
and self energies $\Sigma_{\sigma\sigma^{\prime}}^{(p)}(\mathbf{k})$ are
\begin{equation}
\Delta_{jl}(\mathbf{k})=-\frac{1}{\Omega}\sum_{i=1}^{4}\sum_{\mathbf{q}%
}V_{\mathbf{q}}^{eh}U_{ji}^{\ast}(\mathbf{k}+\mathbf{q})U_{li}(\mathbf{k}%
+\mathbf{q})f(E_{i}(\mathbf{k}+\mathbf{q})),\label{formula2}%
\end{equation}%
\begin{align}
\Sigma_{jl}^{(e)}(\mathbf{k}) &  =\frac{1}{\Omega}\sum_{i=1}^{4}%
\sum_{\mathbf{q}}V_{\mathbf{q}}^{ee}U_{ji}^{\ast}(\mathbf{k}+\mathbf{q}%
)U_{li}(\mathbf{k}+\mathbf{q})f(E_{i}(\mathbf{k}+\mathbf{q})),\label{formula3}%
\\
\Sigma_{jl}^{(h)}(\mathbf{k}) &  =\frac{1}{\Omega}\sum_{i=1}^{4}%
\sum_{\mathbf{q}}V_{\mathbf{q}}^{hh}U_{ji}^{\ast}(\mathbf{k}+\mathbf{q}%
)U_{li}(\mathbf{k}+\mathbf{q})f(E_{i}(\mathbf{k}+\mathbf{q})),\label{formula4}%
\end{align}%
\begin{align}
n_{e} &  =\frac{1}{\Omega}\sum_{i=1}^{2}\sum_{j=2}^{3}\sum_{\mathbf{k}%
}\left\vert U_{ij}(\mathbf{k})\right\vert ^{2}f(E_{j}(\mathbf{k}%
)),\label{formula5}\\
n_{h} &  =\frac{1}{\Omega}\sum_{i=3}^{4}\sum_{j=2}^{3}\sum_{\mathbf{k}}\left[
1-\left\vert U_{ij}(\mathbf{k})\right\vert ^{2}f(E_{j}(\mathbf{k}))\right]
,\label{formula6}%
\end{align}
where $f\left(  E_{i}(\mathbf{k})\right)  $=$1/(1+e^{E_{i}(\mathbf{k})/k_{B}%
T})$ is the Fermi distribution function and $E_{i}(\mathbf{k})$ ($i=1,...,4$)
are the eigen-energies of $H_{MF}(\mathbf{k})$. In Table I we give an explicit
correspondence between $\Delta_{\sigma\sigma^{\prime}}(\mathbf{k})$,
$\Sigma_{\sigma\sigma^{\prime}}^{(p)}(\mathbf{k})$ and $\Delta_{jl}%
(\mathbf{k})$, $\Sigma_{jl}^{(p)}(\mathbf{k})$.%

\[
\overset{\text{TABLE I. The correspondence between }\Delta_{\sigma
\sigma^{\prime}}(\mathbf{k})\text{, }\Sigma_{\sigma\sigma^{\prime}}%
^{(p)}(\mathbf{k})\text{ and }\Delta_{jl}(\mathbf{k})\text{, }\Sigma
_{jl}^{(p)}(\mathbf{k})}{%
\begin{tabular}
[c]{|cc|cc|cc|}\hline\hline
$\Delta_{\sigma\sigma^{\prime}}(\mathbf{k})$ & $\Delta_{jl}(\mathbf{k})$ &
$\Sigma_{\sigma\sigma^{\prime}}^{(e)}(\mathbf{k})$ & $\Sigma_{jl}%
^{(e)}(\mathbf{k})$ & $\Sigma_{\sigma\sigma^{\prime}}^{(h)}(\mathbf{k})$ &
$\Sigma_{jl}^{(h)}(\mathbf{k})$\\\hline
$\sigma\sigma^{\prime}$ & $jl$ & $\sigma\sigma^{\prime}$ & $jl$ &
$\sigma\sigma^{\prime}$ & $jl$\\\hline
$\uparrow\uparrow$ & 13 & $\uparrow\uparrow$ & 11 & $\uparrow\uparrow$ & 33\\
$\uparrow\downarrow$ & 14 & $\uparrow\downarrow$ & 12 & $\uparrow\downarrow$ &
34\\
$\downarrow\uparrow$ & 23 & $\downarrow\uparrow$ & 21 & $\downarrow\uparrow$ &
43\\
$\downarrow\downarrow$ & 24 & $\downarrow\downarrow$ & 22 & $\downarrow
\downarrow$ & 44\\\hline
\end{tabular}
\ \ }%
\]
In addition, for the present 2D case the average interparticle spacing is
given by \cite{Strinati}
\begin{equation}
r_{s}=\frac{1}{\sqrt{\frac{\pi}{2}\left(  n_{e}+n_{h}\right)  }}.
\label{formula7}%
\end{equation}

\begin{figure}[ptb]
\begin{center}
\includegraphics[width=0.5\linewidth]{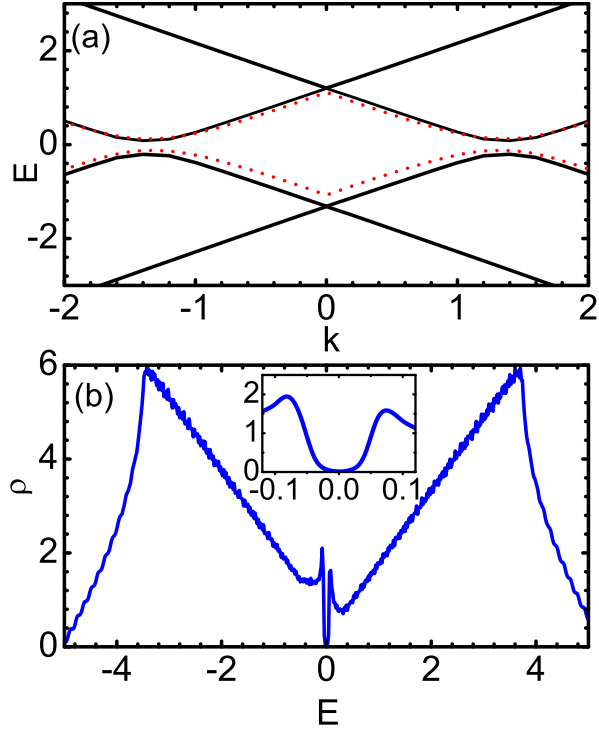}
\end{center}
\caption{(Color online) (a) Exciton's energy spectrum with $r_{s}$=$1$.$5$ and
$\alpha$=$0$. The solid and dashed lines correspond to Eqs. (\ref{formula1})
and (\ref{formula8}), respectively; (b) Exciton density of the states with
$r_{s}$=$5$ and $\alpha$=$0$. }%
\label{f4}%
\end{figure}

\begin{figure}[ptbptb]
\begin{center}
\includegraphics[width=0.6\linewidth]{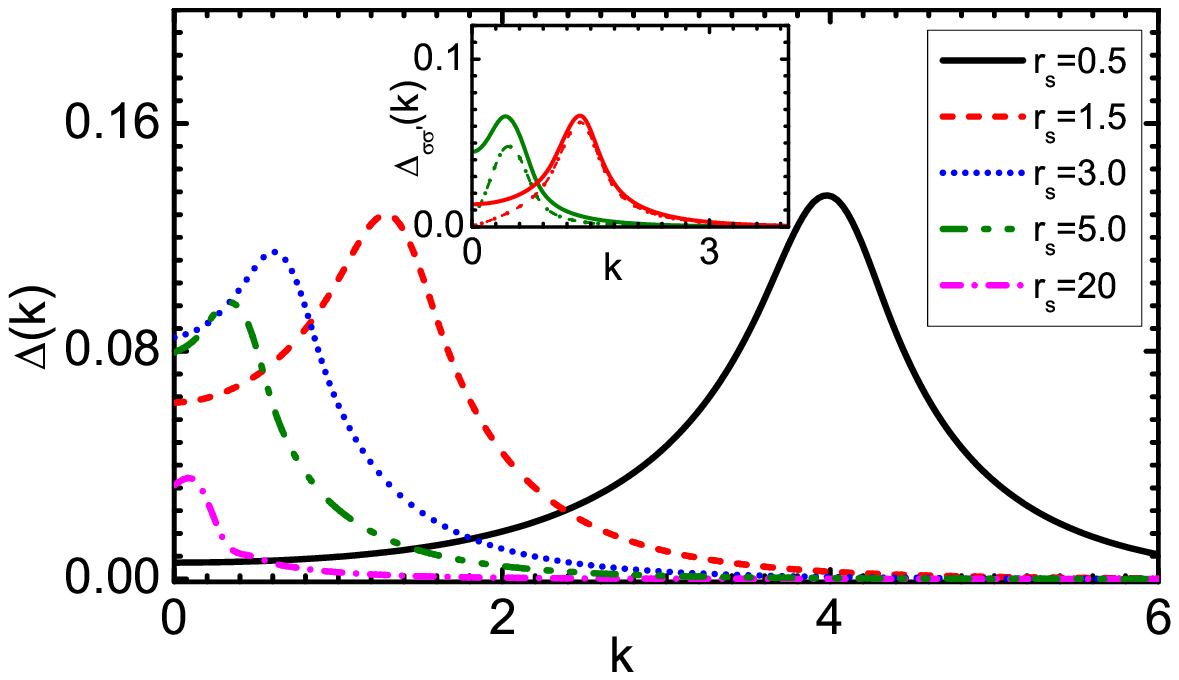}
\end{center}
\caption{(Color online) Wave-vector dependence of the gap function
$\Delta(\mathbf{k})$ for $\alpha$=$0$ and several values of $r_{s}$. Inset:
calculated $\Delta_{\sigma\sigma^{\prime}}(\mathbf{k})$ from original
Hamiltonian (\ref{formula1}) at $r_{s}$=$1.5$ and $5.0$. The solid and dashed
lines are corresponding to $\Delta_{\uparrow\uparrow}$ (=$\Delta
_{\downarrow\downarrow}$) and $\Delta_{\uparrow\downarrow}$ (=$\Delta
_{\downarrow\uparrow}$), respectively. Comparing with $\Delta_{\sigma
\sigma^{\prime}}(\mathbf{k})$ in the inset, we can approximately use
$\Delta(\mathbf{k})$ replacing $\Delta_{\sigma\sigma^{\prime}}(\mathbf{k})$ to
study the general properties of the order parameters without other
spin-dependent interactions.}%
\label{f2}%
\end{figure}

Many meaningful physical quantities, including the order parameters, can be
obtained by self-consistently solving four-band Eqs. (\ref{formula2}%
)-(\ref{formula6}) with the confinement of the electron and hole number
densities. We numerically calculate the exciton's energy spectrum and the
order parameters under different exciton number densities: $r_{s}$=$1.5$,
$\alpha$=$0$ and $5.0$, $\alpha$=$0$. Here the density imbalance parameter
$\alpha$ is defined as $\alpha\mathtt{\equiv}\left(  n_{e}\mathtt{-}%
n_{h}\right)  /\left(  n_{e}\mathtt{+}n_{h}\right)  $. The calculated results
are correspondingly shown by solid lines in Fig. \ref{f4}(a) and the inset in
Fig. \ref{f2}.

Because the main goal of this paper is to focus on the general properties of
the order parameters and neglect the other spin-dependent physical conditions,
such as the effect of the Rashba-type spin-orbit coupling by surface inversion
asymmetry, we plan to simplify our TIB model, i.e., to define a single order
parameter $\Delta(\mathbf{k})$, which can approximately replace the four
$\Delta_{\sigma\sigma^{\prime}}(\mathbf{k})$. Similar to that in the
semiconductor case \cite{Strinati}, the corresponding simplified
grand-canonical Hamiltonian describing this TIB system can be approximately
written as%
\begin{align}
H  &  =\sum_{\mathbf{k},p}\left(  \epsilon_{\mathbf{k}p}-\mu_{p}\right)
c_{\mathbf{k}p}^{\dag}c_{\mathbf{k}p}+\frac{1}{2\Omega}\label{formula8}\\
&  \times\sum_{\substack{\mathbf{k},\mathbf{k}^{\prime},\mathbf{q}%
\\p,p^{\prime}}}V_{\mathbf{k}-\mathbf{k}^{\prime}}^{pp^{\prime}}%
c_{\mathbf{k}+\mathbf{q}/2p}^{\dag}c_{-\mathbf{k}+\mathbf{q}/2p^{\prime}%
}^{\dag}c_{-\mathbf{k}^{\prime}+\mathbf{q}/2p^{\prime}}c_{\mathbf{k}^{\prime
}+\mathbf{q}/2p}.\nonumber
\end{align}
With the SCMF theory, Eq. (\ref{formula8}) can be rewritten in a
$2\mathtt{\times}2$ matrix in the basis $(e,h)^{T}$, the relevant mean-field
equations to be solved for the variables $\mu_{e}$, $\mu_{h}$, and the gap
function $\Delta_{\mathbf{k}}$ are
\begin{equation}
\Delta_{\mathbf{k}}=-\frac{1}{\Omega}\sum_{\mathbf{k}^{\prime}}V_{\mathbf{k}%
-\mathbf{k}^{\prime}}^{eh}\frac{\Delta_{\mathbf{k}^{\prime}}}{2E_{\mathbf{k}%
^{\prime}}}\left[  f(E_{\mathbf{k}^{\prime}}^{+})-f(E_{\mathbf{k}^{\prime}%
}^{-})\right]  , \label{formula9}%
\end{equation}%
\begin{align}
\Sigma_{\mathbf{k}}^{e}  &  =\frac{1}{\Omega}\sum_{\mathbf{k}^{\prime}%
}V_{\mathbf{k}-\mathbf{k}^{\prime}}^{ee}\left[  u_{\mathbf{k}}^{2}%
f(E_{\mathbf{k}^{\prime}}^{+})+v_{\mathbf{k}}^{2}f(E_{\mathbf{k}^{\prime}}%
^{-})\right]  ,\label{formula10}\\
\Sigma_{\mathbf{k}}^{h}  &  =\frac{1}{\Omega}\sum_{\mathbf{k}^{\prime}%
}V_{\mathbf{k}-\mathbf{k}^{\prime}}^{hh}\left[  v_{\mathbf{k}}^{2}%
f(E_{\mathbf{k}^{\prime}}^{+})+u_{\mathbf{k}}^{2}f(E_{\mathbf{k}^{\prime}}%
^{-})\right]  , \label{formula11}%
\end{align}%
\begin{align}
n_{e}  &  =\frac{1}{\Omega}\sum_{\mathbf{k}}\left\{  u_{\mathbf{k}}%
^{2}f(E_{\mathbf{k}}^{+})+v_{\mathbf{k}}^{2}\left[  f(E_{\mathbf{k}}%
^{-})\right]  \right\}  ,\label{formula12}\\
n_{h}  &  =\frac{1}{\Omega}\sum_{\mathbf{k}}\left\{  u_{\mathbf{k}}%
^{2}[1-f(E_{\mathbf{k}}^{-})\}+v_{\mathbf{k}}^{2}\left[  1-f(E_{\mathbf{k}%
}^{+})\right]  \right\}  , \label{formula13}%
\end{align}
where $u_{\mathbf{k}}^{2}$=$1\mathtt{-}v_{\mathbf{k}}^{2}$=$\frac{1}{2}\left(
1\mathtt{+}\xi_{\mathbf{k}}/E_{\mathbf{k}}\right)  $, and $E_{\mathbf{k}}%
^{\pm}$=$\delta\xi_{\mathbf{k}}\mathtt{\pm}E_{\mathbf{k}}$ with $\delta
\xi_{\mathbf{k}}$=$\frac{1}{2}\left(  \xi_{\mathbf{k}e}\mathtt{+}%
\xi_{\mathbf{k}h}\right)  $ and $E_{\mathbf{k}}$=$\sqrt{\xi_{\mathbf{k}}%
^{2}\mathtt{+}\Delta_{\mathbf{k}}^{2}}$ that are given by $\xi_{\mathbf{k}p}%
$=$\epsilon_{\mathbf{k}p}\mathtt{-}\mu_{p}\mathtt{+}\Sigma_{\mathbf{k}}^{p}$
($p$=$e,h$). 

We also self-consistently calculate the exciton's energy spectrum from
two-band Eqs. (\ref{formula9})-(\ref{formula13}). The result, for comparison
with the original exact four-band results from Eqs. (\ref{formula2}%
)-(\ref{formula6}), is plotted in Fig. \ref{f4}(a) with red dashed lines under
the same parameters as used in four-band calculations. The corresponding
density of states is shown in Fig. \ref{f4}(b). From Fig. \ref{f4}(a) one can
clearly find that the exciton energy spectrum within the two-band
approximation is wonderfully consistent with that within the exct four-band
formalism. Another character found from Figs. \ref{f4}(a) and \ref{f4}(b) is
that there is an evident stable energy gap protecting the excitonic
condensation. In addition, we would like to point out that the parity of the
linear dispersion relations of the particles and holes is odd, while the
parity of the particle-particle and hole-hole Coulomb interaction is even.
This parity asymmetry results in the energy-shift in Fig. \ref{f4}(a) and the
corresponding DOS asymmetry in Fig. \ref{f4}(b) as well as the asymmetry in
Fig. \ref{f3a} below. In the following of this paper, all the numerical
results except for those shown in the inset of Fig. 3 are calculated from
two-band SCMF Eqs. (\ref{formula9})-(\ref{formula13}).

\section{Numerical results and application to the Bi$_{2}$Se$_{3}$-family
material}

First, we calculate the wave-vector dependence of $\Delta(\mathbf{k})$ for
equal densities ($\alpha$=$0$) and several values of $r_{s}$. The results are
shown in Fig. \ref{f2}. We can find the generic feature of the BCS-BEC
crossover behavior similar to that in the semiconductor bilayers. However, the
striking character in the TI bilayers is that the maximum value of
$\Delta(\mathbf{k})$ in the BCS limit is much larger than that in the
traditional semiconductor electron-hole bilayers \cite{Strinati}. This
prominent difference means that the BCS phase of TIB is more robust than that
of the semiconductor bilayer for equal-density case. Also shown in Fig. 3
(inset) are the calculated four-band gap functions $\Delta_{\sigma
\sigma^{\prime}}(\mathbf{k})$ at $r_{s}$=1.5 and 5.0, which show the
approximate coincidence in amplitude with the two-band result of
$\Delta(\mathbf{k})$.

\begin{figure}[ptb]
\begin{center}
\includegraphics[width=0.6\linewidth]{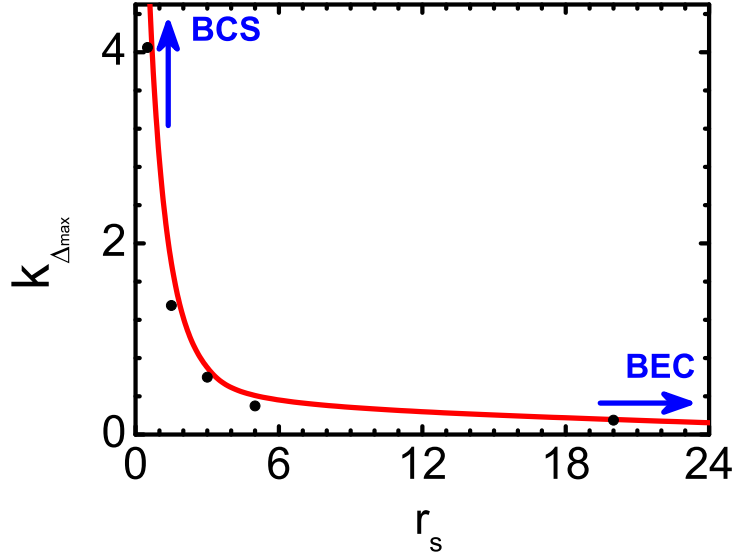}
\end{center}
\caption{(Color online) BCS-BEC phase transform: The momentum magnitude $k$ at
which the order parameter takes its maximum value $\Delta_{\max}$ versus
$r_{s}$ (or number density) at $\alpha$=$0$. The dots are the calculated data,
while the solid line is to guide the eyes.}%
\label{f33}%
\end{figure}

Because there are no obvious interface between BCS and BEC regimes in terms of
the density, we plot in Fig. \ref{f33} the calculated momentum magnitude $k$
at which the order parameter takes its maximum value $\Delta_{\max}$ versus
$r_{s}$ at $\alpha$=$0$. From this figure, one can see that as $r_{s}%
\mathtt{\longrightarrow}0$, the number density $n_{e}$ ($n_{h}$) and
$k_{\Delta_{\max}}$ tend to infinity, the exciton's phase is in the BCS
regime. On the other hand, as $r_{s}\mathtt{\longrightarrow}\infty$, the
number density $n_{e}$ ($n_{h}$) and $k_{\Delta_{\max}}$ tend to $0$, and the
exciton's phase is now in the BEC regime. As $r_{s}$ takes a moderate value,
the system is in a mixed regime. 

\begin{figure}[ptb]
\begin{center}
\includegraphics[width=0.6\linewidth]{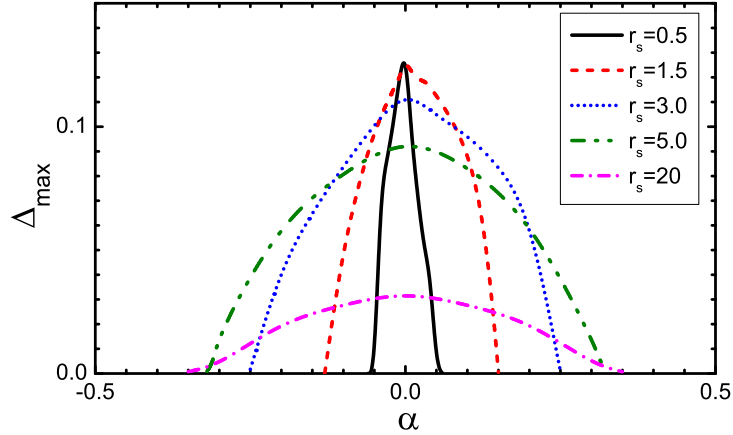}
\end{center}
\caption{(Color online) Maximum value $\Delta_{\max}$=$\max\left\{
\Delta_{\mathbf{k}}\right\}  $ as a function of $\alpha$ for $d$=$1$ and
several values of $r_{s}$. }%
\label{f3a}%
\end{figure}

The effect of $\alpha$ on $\Delta_{\max}$ is shown in Fig. \ref{f3a}, where
$\Delta_{\max}$=$\max\left\{  \Delta_{\mathbf{k}}\right\}  $. It is evident to
find that the density imbalance actually suppresses $\Delta_{\max}$ and it has
different effects on two sides of the crossover. In the BEC regime, the main
effect of the density imbalance is to reduce the number of electron-hole
pairs, which results in that the superfluid properties are less sensitive to
density imbalance. In the BCS regime, the density imbalance leads to the
mismatch of the Fermi surfaces of electrons and holes and the finite momentum
pairing, which is easier to be broken. However, comparing with that in the
traditional semiconductor bilayers, we find that the superfluid property in
the BEC phase in our case is more sensitive to electron-hole density
imbalance. As an example, for $r_{s}$=$20$ the maximum of gap function
$\Delta_{\max}$ for TIB disappears as $\alpha$ takes a value smaller than
$0.5$, while it always takes finite values at $\alpha$ varies in the whole
zone $\left(  -1,1\right)  $ for the traditional semiconductor electron-hole
bilayers \cite{Strinati}.

Now we apply this TIB model to study the condensation of electron-hole pairs
for the topological surface states of the Bi$_{2}$Se$_{3}$-family material.
The two TI films in the left panel of Fig. \ref{f1} now are two ultrathin TI
Bi$_{2}$Se$_{3}$-family films \cite{Hasegawa} (about $80$ \AA \ thick). With
the adopted experimental \cite{Nakajima} lattice constants $a$=$4.143$
\AA \ and $c$=$28.636$ \AA , we calculate the first-principles surface band
structure of Bi$_{2}$Se$_{3}$-family \cite{Wang} by a simple supercell
approach with spin-orbit coupling included and obtain the approximate
Hamiltonian form describing the gapless surface states of Bi$_{2}$Se$_{3}%
$-family as follows:%
\begin{equation}
H(\mathbf{k})=\gamma k^{2}+\hslash v_{F}\left(  k_{x}\sigma_{y}-k_{y}%
\sigma_{x}\right)  . \label{formula14}%
\end{equation}
Although this Hamiltonian has the same form as that of the conventional
two-dimensional electron gas (2DEG) system with Rashba spin-orbit coupling,
the intrinsic difference between these two kinds of systems is that the
$k$-linear spin-orbit interaction is primary to the TI surface states, while
the parabolic term is dominant in the conventional 2DEG. By fitting the
first-principles results, the parameters in Eq. (\ref{formula14}) are given as
$\gamma$=$0.21$ eV nm$^{2}$ and $\hslash v_{F}$=$0.2$ eV nm (namely, $v_{F}%
$=$3.04\times10^{5}$m/s). That means the energy dispersion around the Dirac
point can be accurately described by $\epsilon_{\mathbf{k}}$=$\pm\hslash
v_{F}|\mathbf{k}|$ when the wave-vector $|\mathbf{k}|$ is much smaller than
$1.0$ nm$^{-1}$. For numerical calculation, we choose nm as the length unit
and $0.2$ eV as the energy unit in the following discussion. The dielectric
constant $\varepsilon$=$1$ and the spacer width $d$=$10$ \AA . In fact, the
condition that the wave-vector $|\mathbf{k}|$ is much smaller than $1.0$
nm$^{-1}$ requires that only for $r_{s}\geq5$, then the TIB model is valid for
Bi$_{2}$Se$_{3}$-family material. This means that the BEC phase is most
possible to emerge in Bi$_{2}$Se$_{3}$-family bilayer system.

\begin{figure}[ptb]
\begin{center}
\includegraphics[width=0.7\linewidth]{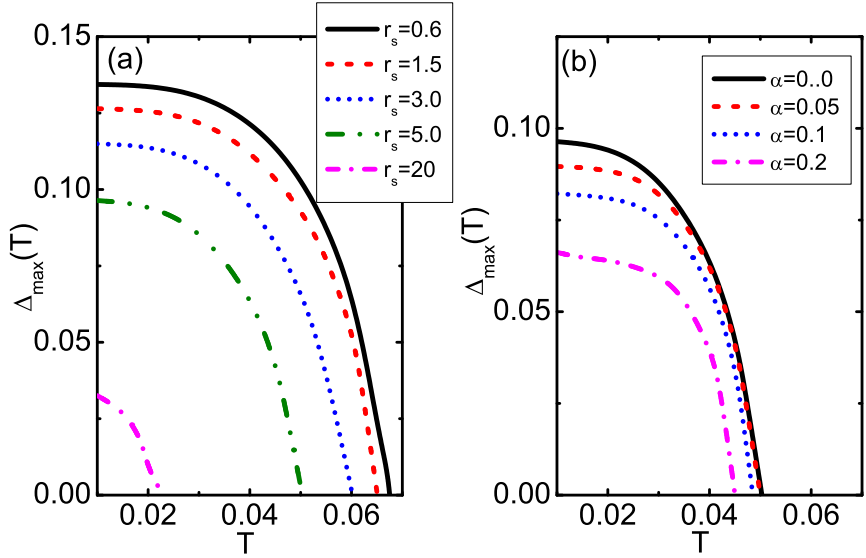}
\end{center}
\caption{(Color online) (a) Maximum value $\Delta_{\max}$=$\max\left\{
\Delta_{\mathbf{k}}\right\}  $ vs temperature $T$ for $d$=$1$, $\alpha$=$0$,
and several values of $r_{s}$. (b) $\Delta_{\max}$ as a function of the
temperature $T$ at $r_{s}$=$5$ and several values of $\alpha$.}%
\label{f5}%
\end{figure}

Now, let us discuss the critical temperature of this TIB system. The relation
between the $\Delta_{\max}$ and temperature $T$ is respectively shown in Fig.
\ref{f5}(a) for $d$=$1$, $\alpha$=$0$, and several values of $r_{s}$, and
\ref{f5}(b) for $d$=$1$, $r_{s}$=$5$.$0$, and several values of $\alpha$. From
Fig. \ref{f5}(a), we can find that the critical temperature $T_{c}$ decreases
as $r_{s}$ increases (i.e., as the particle density decreases). For the
Bi$_{2}$Se$_{3}$-family bilayer at $r_{s}$=$5$.$0$, the critical temperature
$T_{c}$ is calculated as 0.05 in unit of 0.2 eV. That means the critical
temperature $T_{c}$ is about $8\mathtt{\sim}10$ meV ($100$ K), which is much
higher than that in the traditional semiconductor electron-hole bilayers.
Although the Bi$_{2}$Se$_{3}$-family TIB system is in the BEC phases ($r_{s}%
$=$5$.$0$, $20.0$), the numerical calculated results shown in Fig. \ref{f5}(a)
are consistent with the general relation of BCS superconductor,%
\begin{equation}
\frac{2\Delta(0)}{T_{c}}=2\pi e^{-\gamma}\approx3.53, \label{formula15}%
\end{equation}
where $\Delta(0)$ is the energy gap at zero temperature. The introduced
electron-hole density imbalance ($\alpha\mathtt{\neq}0$) can reduce the
critical temperature. This character is clearly shown in Fig. \ref{f5}(b): by
increasing the density imbalance $\alpha$, the critical temperature $T_{c}$
decreases. 

\begin{figure}[ptb]
\begin{center}
\includegraphics[width=0.5\linewidth]{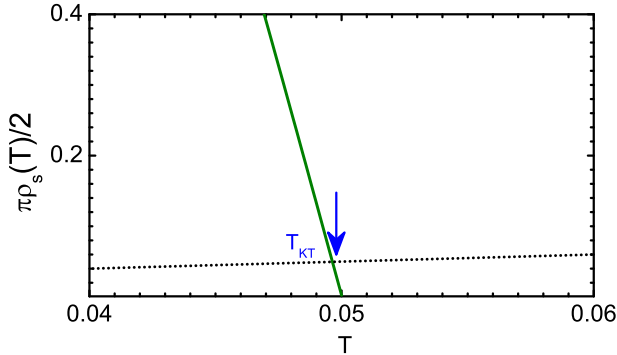}
\end{center}
\caption{(Color online) The calculated $T_{KT}$ at $r_{s}$=$5$ and $\alpha
$=$0$.}%
\label{f6}%
\end{figure}

As it is known that in 2D superfluids, the critical temperature is often
substantially overestimated by mean-filed theory. It is ultimately limited by
entropically driven vortex and antivortex proliferation at the
Kosterlitz-Thouless (KT) transition temperature $T_{\text{KT}}$=$\frac{\pi}%
{2}\rho_{s}(T_{\text{KT}})$ with $\rho_{s}(T)$ being the superfluid density
(the phase stiffness). Ref. \cite{Mac} gives an approximate formula to
calculate the counterflow current, which is read as
\begin{equation}
\rho_{s}(T)\approx\frac{v^{2}\hslash^{2}}{16\pi T}\int kdk\left[  \sec
\text{h}^{2}\left(  \frac{\Delta^{z}}{2T}\right)  -\sec\text{h}^{2}\left(
\frac{\Delta}{2T}\right)  \right]  , \label{formula16}%
\end{equation}
where $\Delta^{z}=\frac{-\mu_{e}+\mu_{h}+\Sigma_{\mathbf{k}}^{e}%
-\Sigma_{\mathbf{k}}^{h}}{2}$, and $\Delta=\sqrt{\left(  \Delta^{z}\right)
^{2}}+\sqrt{\Delta_{\mathbf{k}}^{2}}$. We adopt this formula to calculate the
superfluid density. The temperature dependence of superfluid density is shown
in Fig. \ref{f6} at $r_{s}$=$5$ and $\alpha$=$0$. From Fig. \ref{f6}, it is
evident to estimate that the KT transition temperature $T_{\text{KT}}$ is
about $0.05$ in unit of $0.2$ eV. Comparing with the critical temperature
$T_{c}$ in Fig. \ref{f5} at $r_{s}$=$5$ and $\alpha=0$, the striking
conclusion is reached: $T_{c}\mathtt{\approx}$ $T_{\text{KT}}$, which means
that high-temperature ($\mathtt{\sim}$100 K) excitonic condensation may occur
in the Bi$_{2}$Se$_{3}$-family TIB system. On the other hand, we can estimate
the KT temperature with the zero-temperature phase stiffness $\rho_{s}%
(T$=$0)\mathtt{\approx}E_{F}/4\pi$ which is similar to the graphene bilayers
\cite{Mac}. Considering the case shown in Fig. \ref{f4}, the Fermi energy
$E_{F}$ can be numerically calculated and is given to be $\mathtt{\sim}0.4$
(in unit of $0.2$ eV). Hence, the KT temperature is estimated as
$T_{\text{KT}}\mathtt{\approx}E_{F}/8\mathtt{\approx}0.05$ in unit of $0.2$
eV. This means that the two estimated methods are consistent and the
high-temperature excitonic condensation can emerge in the Bi$_{2}$Se$_{3}%
$-family TIB system. 

\section{Summary and conclusions}

In summary, we have performed a generic TIB model to study the excitonic
condensation with the SCMF theory for the topological surface states. Similar
to the traditional semiconductor electron-hole bilayers, the TIB system
presents the crossover behavior from BCS limit to BEC limit by changing the
exciton's density. However, two prominent novel characters different from the
traditional semiconductor electron-hole bilayers are found. One is that the
superfluid property in the BEC phase is more sensitive to electron-hole
density imbalance. The other is that the BCS phase is more robust than that of
the semiconductor bilayer. Applying this TIB model to Bi$_{2}$Se$_{3}$-family
material, we find that the BEC phase is most possibly observed in experiment.
Moveover, we theoretically estimate the critical temperature for the Bi$_{2}%
$Se$_{3}$-family TIB system and find that it is much higher than that in the
traditional semiconductor electron-hole bilayers. For example, at $r_{s}$=$5$
and $\alpha$=$0$, the critical temperature $T_{c}$ is obtained as about $100$
K. We have also studied the phase stiffness and find that the KT transition
doesn't suppress the critical temperature for Bi$_{2}$Se$_{3}$-family in SCMF approximation.

\begin{acknowledgments}
This work was supported by NSFC under Grants No. 90921003 and No. 10904005,
and by the National Basic Research Program of China (973 Program) under Grant
No. 2009CB929103.
\end{acknowledgments}


\end{document}